\documentstyle{article}
\begin{document}

\title{Transmit Classical and Quantum Information Secretly }
\author{{Li Yang(1) and Ling-An Wu(2)}\\
 {\small (1)State Key Laboratory of Information Security,}\\
   {\small Graduate School of
   Chinese Academy of Sciences,}
   {\small Beijing 100039,P. R. of China }\\
   {\small(2)Institute of Physics, Chinese Academy of Sciences,
    Beijing 100086, China}}

\date{}
\maketitle

\begin{minipage}{120mm}
\vskip 0.3in
                       $ \mathbf{Abstract}$
    This note presents a practical quantum cryptography protocol for
transmitting classical and quantum information secretly and
directly.
\end{minipage}

\vskip 0.5in

     First, let's consider the transmission of classical message. The protocol is:\\

      1. Alice prepares n bits for transmission:
\begin{equation}\label{1}
  a_{1}a_{2}\cdots a_{n},
\end{equation}
and chooses
\begin{equation}\label{2}
  \{\varphi_{Ai}|i=1,2,\cdots,n\}
\end{equation}
randomly from a K-element set
\begin{equation}\label{3}
  \{\alpha_{k}=\frac{k\pi}{K}, k=0,1,\cdots,K-1\}
\end{equation}
by means of local random number sources.\\

      2. Alice prepares n single-photons, the i-th be in the state
\begin{equation}\label{4}
  |\Psi_{i}\rangle_{A0}=(a_{i}\oplus1)|H\rangle+a_{i}|V\rangle,
\end{equation}
where the $\oplus$ is the plus in $F_{2}$; then changes the
polarization directions of photons separately and get
\begin{equation}\label{5}
  |\Psi_{i}\rangle_{A1}=(a_{i}\oplus1)|\varphi_{Ai}\rangle+a_{i}|\varphi_{Ai}+\frac{\pi}{2}\rangle;
\end{equation}
then sends these photons to Bob one by one.\\

      3. Bob chooses
\begin{equation}\label{6}
  \{\varphi_{Bi}|i=1,2,\cdots,n\}
\end{equation}
randomly independently from the K-element set (3) by means of
local random number source, and changes the polarization
directions of photons separately as below:
\begin{equation}\label{7}
  |\Psi_{i}\rangle_{B1}=(a_{i}\oplus1)|\varphi_{Ai}+\varphi_{Bi}\rangle
  +a_{i}|\varphi_{Ai}+\varphi_{Bi}+\frac{\pi}{2}\rangle;
\end{equation}
then Bob sends back these photons to Alice.\\

      4. Alice changes the polarization direction of the photons
      again and gets:
\begin{equation}\label{8}
  |\Psi_{i}\rangle_{A2}=(a_{i}\oplus1)|\varphi_{Bi}\rangle
  +a_{i}|\varphi_{Bi}+\frac{\pi}{2}\rangle;
\end{equation}
then sends them to Bob again.\\

      5. Bob changes the polarization direction of the photons again and gets:
\begin{equation}\label{9}
  |\Psi_{i}\rangle_{B2}=(a_{i}\oplus1)|H\rangle+a_{i}|V\rangle;
\end{equation}
then measures the photons in bases $\{|H\rangle,|V\rangle\}$ one
by one and gets the message (1).\\

    Now let's consider the secret transmission of quantum information.
It is obviously that the changes needed are trivial. The protocol
becomes:\\

      1. Alice chooses n qubits for transmission:
\begin{equation}\label{1}
  |\varphi_{1}\rangle,|\varphi_{2}\rangle,\cdots ,|\varphi_{n}\rangle,
\end{equation}
and chooses
\begin{equation}\label{2}
  \{\varphi_{Ai}|i=1,2,\cdots,n\}
\end{equation}
randomly from a K-element set
\begin{equation}\label{3}
  \{\alpha_{k}=\frac{k\pi}{K}, k=0,1,\cdots,K-1\}
\end{equation}
by means of local random number sources.\\

      2. Alice prepares n single-photons, the i-th be in the state
\begin{equation}\label{5}
  |\Psi_{i}\rangle_{A1}=|\varphi_{i}+\varphi_{Ai}\rangle;
\end{equation}
sends these photons to Bob one by one.\\

      3. Bob chooses
\begin{equation}\label{6}
  \{\varphi_{Bi}|i=1,2,\cdots,n\}
\end{equation}
randomly from the K-element set (3) by means of local random
number source, and changes the polarization directions of photons
separately as below:
\begin{equation}\label{7}
  |\Psi_{i}\rangle_{B1}=|\varphi_{i}+\varphi_{Ai}+\varphi_{Bi}\rangle;
\end{equation}
then sends back these photons to Alice.\\

      4. Alice changes the polarization direction of the photons
      again and gets:
\begin{equation}\label{8}
  |\Psi_{i}\rangle_{A2}=|\varphi_{i}+\varphi_{Bi}\rangle;
\end{equation}
then sends them to Bob again.\\

      5. Bob changes the polarization direction of the photons again and gets:
\begin{equation}\label{9}
  |\Psi_{i}\rangle_{B2}=|\varphi_{i}\rangle;
\end{equation}
then he  gets the message (10).\\

  Because $\varphi_{Ai},\varphi_{Bi}$are chosen from set (3)
  randomly and independently, Eve cannot get any information from
  simple intercept/resend attack. Unfortunately, These two protocols cannot defend
  man in the middle(of quantum channel only) attack, even though there
  is an authenticated classical channel. To overcome this problem,
  Alice and Bob need to share $\{\varphi_{Ci},i=1,\cdots,n\}$secretly
  before communication. Then, For example, the second protocol becomes:\\

      1. Alice chooses n qubits for transmission:
\begin{equation}\label{1}
  |\varphi_{1}\rangle,|\varphi_{2}\rangle,\cdots ,|\varphi_{n}\rangle,
\end{equation}
and chooses
\begin{equation}\label{2}
  \{\varphi_{Ai}|i=1,2,\cdots,n\}
\end{equation}
randomly from a K-element set
\begin{equation}\label{3}
  \{\alpha_{k}=\frac{k\pi}{K}, k=0,1,\cdots,K-1\}
\end{equation}
by means of local random number sources.\\

      2. Alice prepares n single-photons, the i-th be in the state
\begin{equation}\label{5}
  |\Psi_{i}\rangle_{A1}=|\varphi_{Ci}+\varphi_{i}+\varphi_{Ai}\rangle;
\end{equation}
then sends these photons to Bob one by one.\\

      3. Bob chooses
\begin{equation}\label{6}
  \{\varphi_{Bi}|i=1,2,\cdots,n\}
\end{equation}
randomly from the K-element set (3) by means of local random
number source, and changes the polarization directions of photons
separately:
\begin{equation}\label{7}
  |\Psi_{i}\rangle_{B1}=|\varphi_{Ci}+\varphi_{i}+\varphi_{Ai}+\varphi_{Bi}\rangle;
\end{equation}
then sends them back to Alice.\\

      4. Alice changes the polarization direction of the photons
      again and gets:
\begin{equation}\label{8}
  |\Psi_{i}\rangle_{A2}=|\varphi_{Ci}+\varphi_{i}+\varphi_{Bi}\rangle;
\end{equation}
then sends them to Bob again.\\

      5. Bob changes the polarization direction of the photons again and gets:
\begin{equation}\label{9}
  |\Psi_{i}\rangle_{B2}=|\varphi_{i}\rangle;
\end{equation}
then he gets the message (18).\\

The authentication information $\{\varphi_{Ci},i=1,\cdots,n\}$ can
be used repeatedly under the protection of continuously changed
local random numbers $\{\varphi_{Ai},\varphi_{Bi},i=1,\cdots,n\}$
.\\

[1]C.H.Bennett, et al.,J.Cryptography(1992)5:3-28

\end{document}